\documentclass{PoS}

\usepackage[english]{babel}
\usepackage{amsmath,amssymb,bm}
\usepackage{url}
\usepackage{fontenc}
\usepackage{times}
\usepackage{mathptmx}
\usepackage{graphicx}

\def\txt#1{\text{#1}} 
\def\vc#1{\vec{#1}} 
\newcommand\mycite[1]{[\ref{#1}]} 
\newcommand\mycites[2]{[\ref{#1}, \ref{#2}]} 

\title{Timing X-ray Pulsars with \\ Application to Spacecraft Navigation}

\ShortTitle{Timing X-ray Pulsars with Application to Spacecraft Navigation}

\author{Mike Georg Bernhardt,\;
        Tobias Prinz,\;
        Werner Becker\\
        Max-Planck-Institut f\"ur extraterrestrische Physik\\ 
        Gie\ss{}enbachstr. 1, 85741 Garching, Germany \\
        E-mail: \email{ m.bernhardt@mpe.mpg.de} \\
        \phantom{E-mail: }\email{ tprinz@mpe.mpg.de}\\
        \phantom{E-mail: }\email{ wbecker@mpe.mpg.de}
        }

\author{Ulrich Walter\\
        Institute of Astronautics, Technische Universit\"at M\"unchen \\
        Boltzmannstr. 15, 85748 Garching, Germany \\
        E-mail: \email{ u.walter@lrt.mw.tum.de} \\
        \quad \\
        \quad
        }

\abstract{Usually, positions of spacecraft on interplanetary or deep space missions are determined by radar tracking from ground stations, a method by which uncertainty increases with distance from Earth. As an alternative, a spacecraft equipped with e.g. an X-ray telescope could determine its position autonomoulsy via onboard analysis of X-ray pulsar signals. In order to find out which pulsars are best suited for this approach and what accuracy can be achieved, we build up a database containing the temporal emission characteristics of the $\sim 60$ X-ray pulsars for which a pulsed radiation has been detected by mid 2010.}

\FullConference{High Time Resolution Astrophysics IV - The Era of Extremely Large Telescopes - HTRA-IV,\\
                May 5-7, 2010\\
                Agios Nikolaos, Crete, Greece}

\begin{document}

\section{Introduction}

The conventional way to obtain data for orbit determination is tracking the spacecraft by ground stations on Earth, e.g. with NASA's Deep Space Network (DSN). This technique yields very accurate range and range-rate data along the Earth-spacecraft line, but due to limited angular resolution large errors can occur in perpendicular directions, resulting in positional errors that grow with distance. Interferometric measurements can augment the angular resolution, thereby achieving positional accuracies in the order of 4 km per AU of distance between Earth and spacecraft \mycite{james2}. Thus, especially -- but not exclusively -- for interplanetary and deep space missions, it is desirable to have an autonomous navigation solution that works independently from ground stations.

Pulsars can be used as natural beacons for navigation by comparing pulse arrival times measured onboard the spacecraft with predicted arrival times at an inertial reference location -- e.g. the barycenter of the solar system. The phase difference between the expected and the measured pulse arrival corresponds to a run-time difference along the line of sight towards the pulsar and hence to a range difference in this direction. Full three-dimensional position information can be deduced from the range information along the pulsar lines-of-sight of at least three different pulsars. In general, this procedure results in multiple solutions, which can be reduced by either constraining the possible spacecraft positions to a finite volume around an initial position assumption, or by observing additional pulsars \mycites{esa2}{sheikh2}.

\section{Data Analysis and Results}

In order to decide which pulsars are best suited for spacecraft navigation, we re-analyzed all pulsar timing data from the X-ray satellites XMM-Newton, Chandra and the ROSSI X-ray Timing Explorer. The individual photon arrival times of each data set were corrected for the orbital motion of the detector around Earth/Sun via transformation to the solar system barycenter and for the orbital motion of the pulsar in case of a binary system. To account for the gradual increase of pulse period the spin frequency was modelled as a Taylor expansion,
\begin{align}
  f(t) = f_0 + \dot{f}_0 \cdot (t-t_0)
             + \tfrac{1}{2} \ddot{f}_0 \cdot (t-t_0)^2  , \label{taylor}
\end{align}
with $f_0$, $\dot{f}_0$, $\ddot{f}_0$ being the rotation frequency and its first and second time derivative at some reference epoch $t_0$, and a pulse number $\Phi$ was assigned  to each individual arrival time. Since the frequency equals the rate of change of pulse number, $f = \txt d \Phi/\txt d t$, integration of \eqref{taylor} yields
\begin{align}
  \Phi(t) = \Phi_0 + f_0 \cdot (t-t_0) + \tfrac{1}{2} \dot{f}_0 \cdot (t-t_0)^2 
             + \tfrac{1}{6} \ddot{f}_0 \cdot (t-t_0)^3  , \label{spin-down}
\end{align}
where $\Phi_0$ denotes the pulse number at $t_0$. 
The pulse numbers are used to generate a mean pulse profile reflecting the temporal emission characteristics of the pulsar: (1) For each photon its phase $\phi := (\Phi\;\text{mod}\;1)$ is computed, i.e. the fraction of full pulse period at which the photon was detected; (2) the domain $[0,1)$ of  $\phi$ is divided into $n$ finite intervals 
$\Delta\phi_i := \big[\frac{i-1}{n}\,,\, \frac{i}{n}\big)$ for $i=1, 2, \dots, n$; (3) finally, the mean pulse profile is given by the histogram of photon counts per phase interval.
The optimal value of $n$ depends on the number of recorded photons and the harmonic content of the pulse profile. It can be computed as follows: For each set of arrival times $t_i$ with $i=1, 2, \dots, N$ and phases $\phi_i :=\phi(t_i)$ we calculated the statistical variable
\begin{align}
  Z_m^2 := \frac{2}{N} \sum\limits_{k=1}^{m} 
            \bigg[ \bigg( \sum\limits_{i=1}^{N} \cos k\phi_i \bigg)^{\!2}
                   + \bigg( \sum\limits_{i=1}^{N} \sin k\phi_i \bigg)^{\!2}
            \bigg] ,
\end{align}
which is a measure for the periodicity of the signal \mycite{buccheri2}, and evaluated the expression
\begin{align}
  H := \max_{1 \le m \le 10 } \big( Z_m^2 - 4m + 4 \big) = Z_M^2 - 4M + 4 
\end{align}
to get the optimal number $M$ of harmonics \mycite{dejaeger2}. Then, the formula
\begin{align}
  n = 2.36 \bigg[\sum\limits_{m=1}^M \tfrac{1}{2}m ^2 \big( Z^2_m - Z^2_{m-1}\big) \bigg]^{\!1/3}
\end{align}
with $Z^2_0:=0 $ yields an estimate for the optimal number of phase intervals \mycite{beckertruemper2}.
Since equation \eqref{taylor} may not allow to compute $f(t)$ for times $t$ outside the validity range of the ephemerides, 
the values of $f_0$, $\dot{f}_0$, $\ddot{f}_0$ have to be known at an epoch sufficiently close to the measured photon arrival times. We used pulsar ephemerides from the ATNF database \mycite{atnf2} for the period folding and produced pulse profiles with high signal-to-noise ratios by superposing phase values from several obvservations of the same pulsar.

\begin{figure}[tb]                                                                             
  \centering                                                                                   
  \includegraphics[scale=0.42]{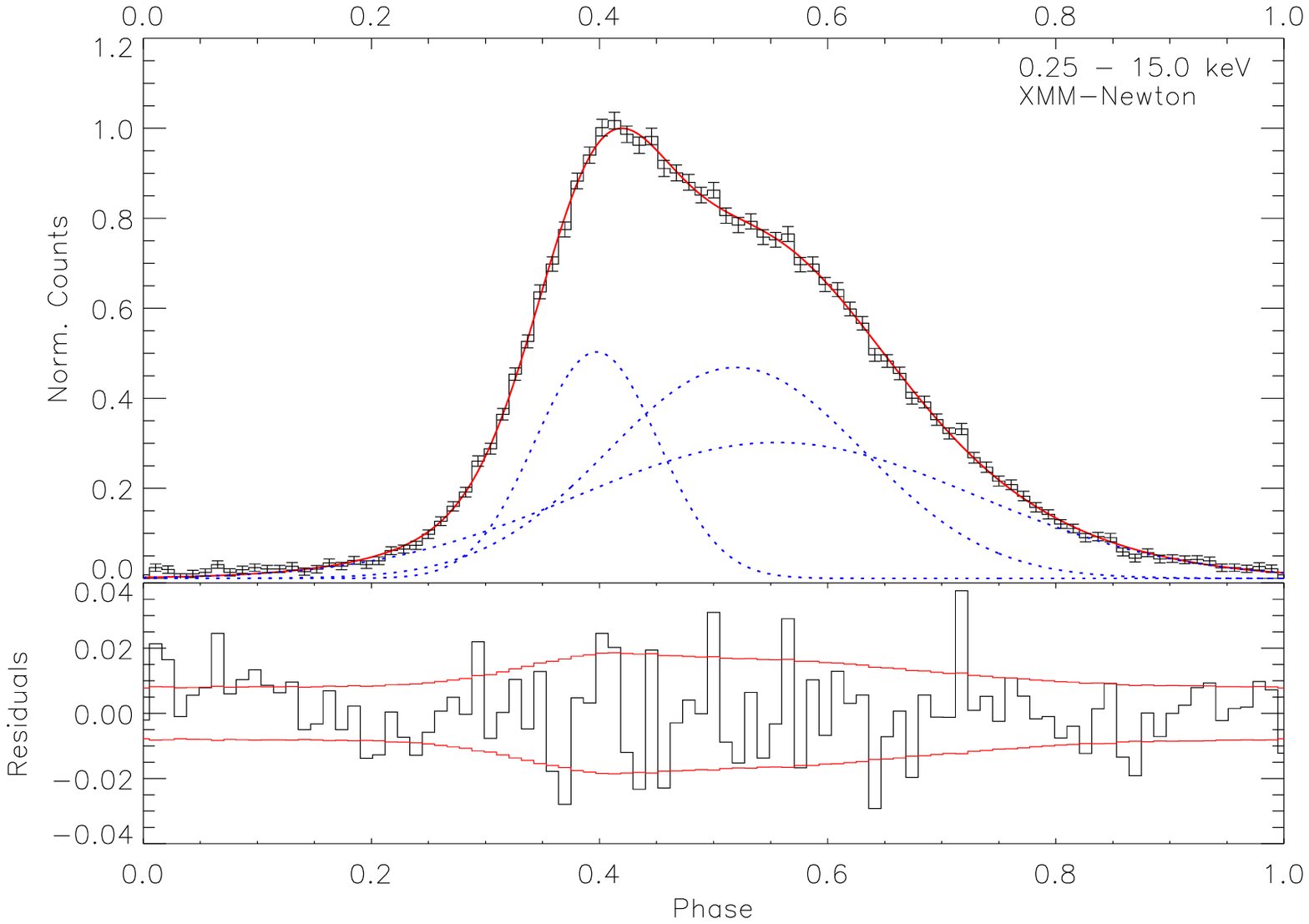}                           
  \includegraphics[scale=0.42]{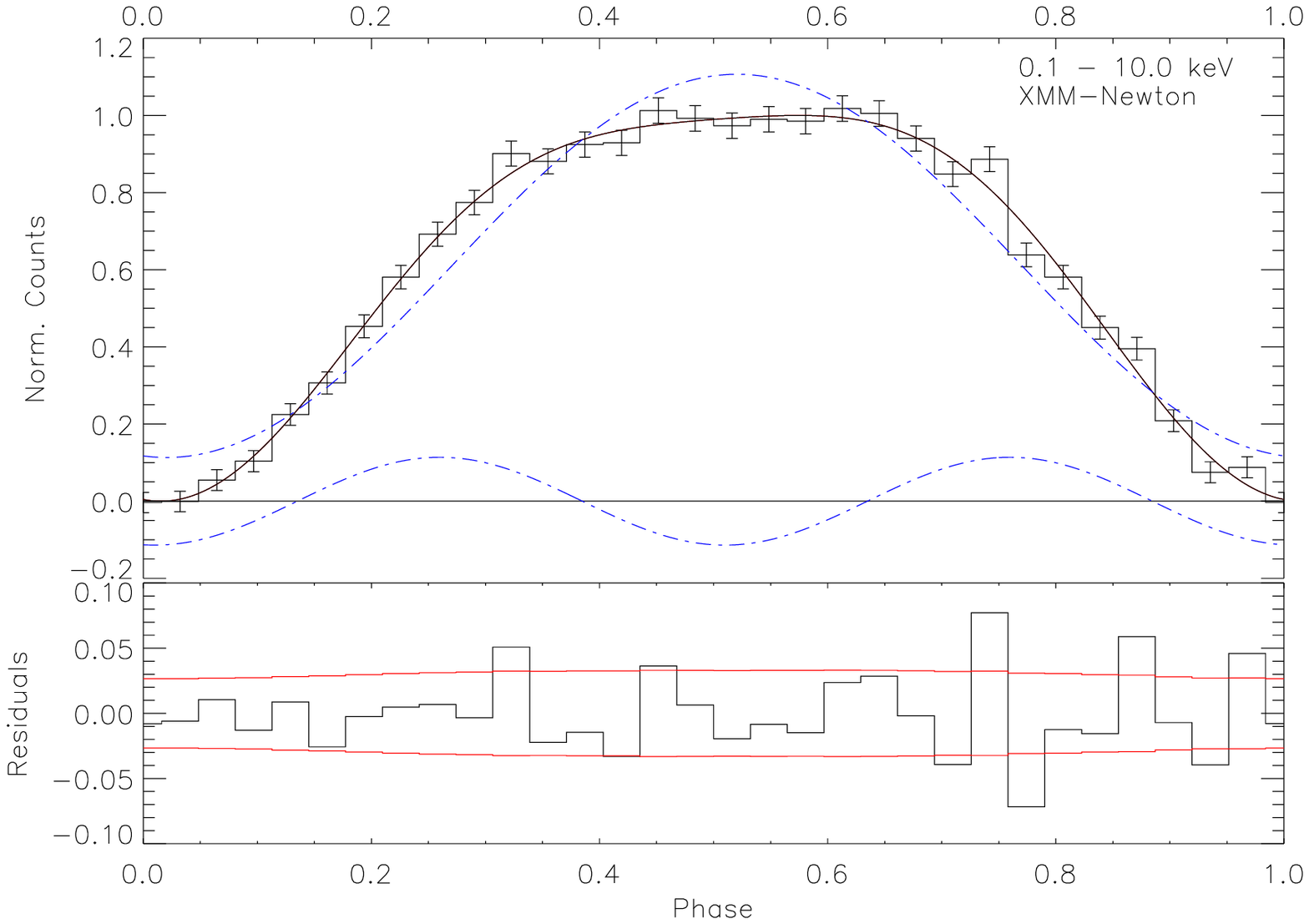}                           
  \caption{Pulse profiles of PSR B1509-58 (left) and PSR J0633+1746 (right).                   
           The data points of the pulse profiles can be approximated by a sum of               
           Gauss or sine functions, respectively.                                              
           The dotted blue lines indicate the individual components of the fit                 
           (three Gaussians for PSR B1509-58 and two sine functions for PSR J0633+1746);       
           their sum is plotted in red.                                                        
           The panels below show the residuals between fit and data points, and the red        
           enveloping lines indicate the 1-$\sigma$ uncertainties of the binned pulse profile.}
  \label{templates1}                                                                           
\end{figure}

The navigation algorithm is based on the comparison of pulse arrivals measured onboard the spacecraft with those measured at the solar system barycenter.%
\footnote{By \emph{pulse arrival} we understand the phase of the global maximum in the pulse profile.} %
For this purpose, representation of the reference profiles by analytical functions is of advantage. Using a least-squares-fit procedure, we found approximations of the profiles by either a sum of Gaussians, 
$\sum_{i} A_i \exp\!\big[\!-\frac{1}{2} \big( \frac{\phi-B_i}{C_i} \big)^2  \big]$,
or of sine functions,
$\sum_{i} A_i \sin ( B_i \phi + C_i ) + D$,
depending on the individual shape of the profile. Concerning the use of Gauss or sine functions, the resulting $M$ value of the $H$-test is a good indicator: For $M\le 2$ the profile is well approximated by a sum of two or three sine functions, whereas for $M\ge 3$ it is often better to use a sum of two or more Gaussians.
Examples of pulse profiles and their analytical representations are shown in Figure~\ref{templates1}.
These pulse profile templates allow us to measure pulse arrivals with high accuracy even for sparse photon statistics -- again by using a least-squares fit of an adequately adjusted template to the profile in question. The typical error of measuring a pulse phase lies in the order of $10^{-3}$, which corresponds to an error in range along the pulsar line-of-sight of about 1~km for the fastest and $30\,000$~km for the slowest rotating pulsar of our database.%
\footnote{PSR B1937+21 with a rotation period $P=1{.}56$~ms and a 1-$\sigma$ error in phase measurement $\delta\phi=2.6\times10^{-3}$; PSR~J1841-0456 with $P=11{.}78$~s and $\delta\phi=9.3\times10^{-3}$. The phase error translates into an error of position along the pulsar line-of-sight according to the formula
$\delta r = cP \delta\phi $, where $c$ is the speed of light.} %
Evidently, the precision of a pulsar based navigation system strongly depends on the choice of pulsars and the accuracy of pulse arrival measurements, wich is subject to the quality of the available templates.

\begin{figure}[hbt]                                                                            
  \centering                                                                                   
  \includegraphics[scale=0.42]{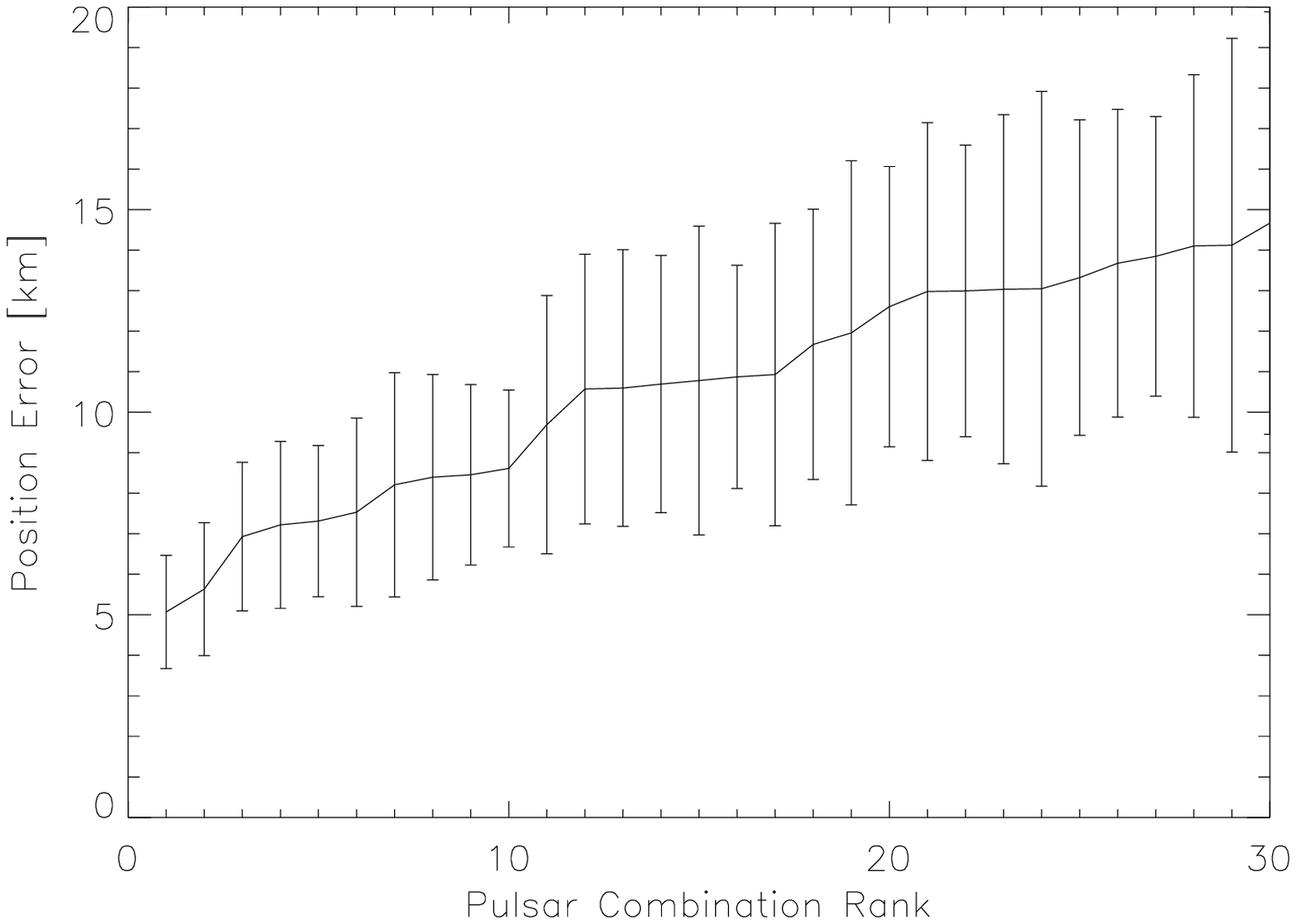}                                       
  \includegraphics[scale=0.42]{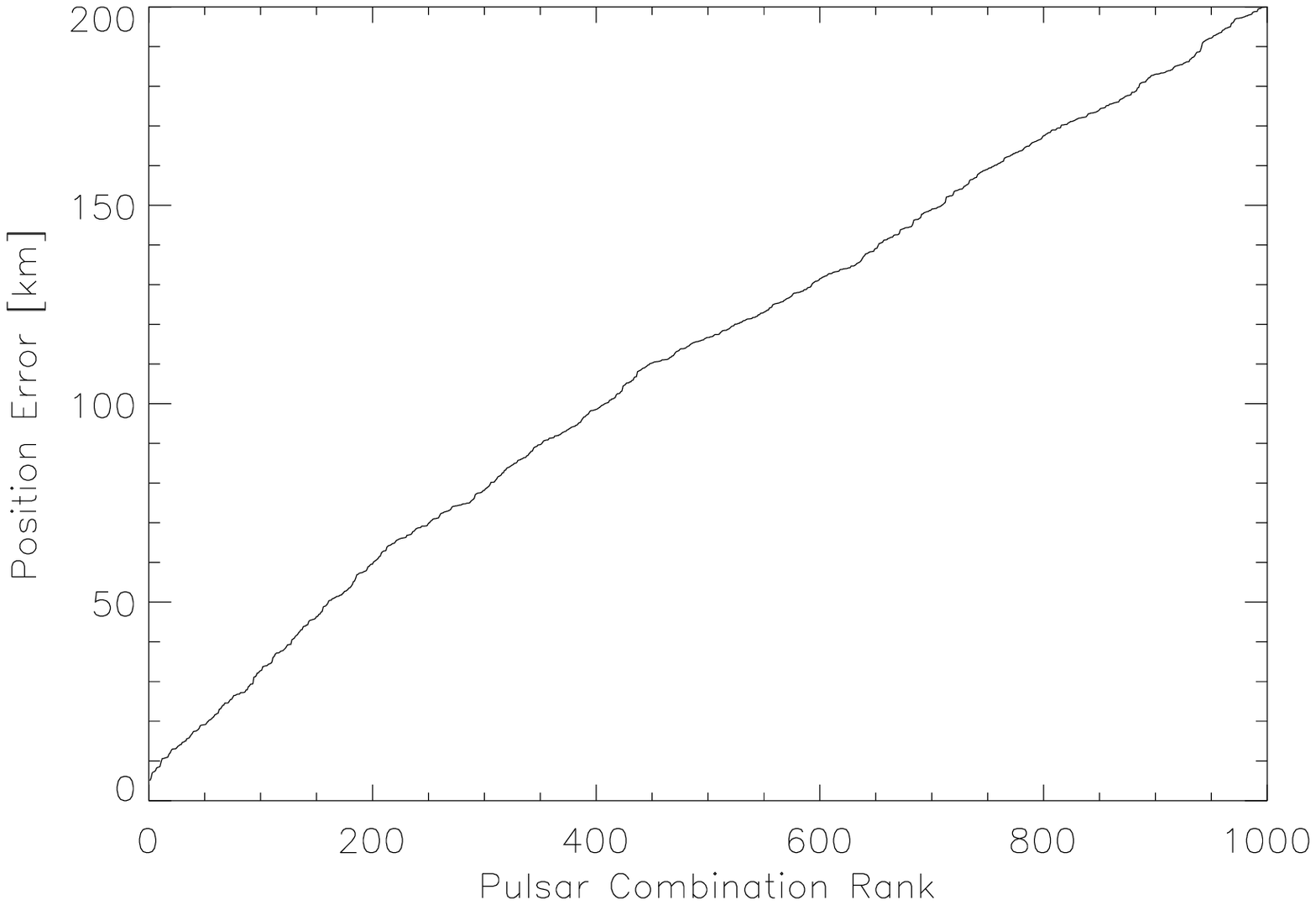}                                     
  \caption{Position errors for various pulsar combinations.                                    
           The diagramme shows mean position errors and standard deviations                    
           for the best 30 combinations (left) and mean position errors for the best 1000      
           combinations (right) of three pulsars each.}                                        
  \label{posdiffs}                                                                             
\end{figure}

As mentioned above, in order to obtain three-dimensional position information, pulsar timing data of at least three different sources have to be collected and analyzed. The spatial arrangement of these pulsars is another, significant parameter of the achievable accuracy. The question of how pulse arrival errors translate into errors of position in space was addressed as follows:
Suppose $\vc r$ to be the position of the detector relative to the solar system barycenter (SSB). A pulse arrival measured at $\vc r$ will in general be shifted by some value $\Delta\phi \in [0,1)$ as compared to a measurement at the SSB. Assuming the pulsar signals to be plane waves propagating through flat spacetime, this phase shift is given by the formula
$\vc n \cdot \vc r = (N + \Delta\phi)cP$,
where $\vc n$ is a unit vector from the SSB towards the pulsar, $N$ is some integer pulse number, $c$ is the speed of light and $P$ is the pulse period at the epoch of measurement.
We chose $\vec r$ to be some fixed point in the solar system and for each pulsar computed 1) the phase shift associated with that location and 2) a sequence of pulse arrival errors, normally distributed around the expected phase shift.
The standard deviations of these normal distributions are known from our study of pulse arrival measurements.
For a combination of three different pulsars the last equation can be written in matrix notation and solved for $\vc r$,
\begin{align}
  \vc r=
  \begin{pmatrix} n_{1x} & n_{1y} & n_{1z}\\ 
                  n_{2x} & n_{2y} & n_{2z}\\
                  n_{3x} & n_{3y} & n_{3z}
  \end{pmatrix}^{\!\!-1} 
  \!
  \begin{pmatrix} (N_1 + \Delta\phi_1)cP_1\\ 
                  (N_2 + \Delta\phi_2)cP_2\\
                  (N_3 + \Delta\phi_3)cP_3
  \end{pmatrix} . \label{position}
\end{align}
Inserting the pulse arrival errors instead of the expected phase shifts in equation \eqref{position} gives sequences of position vectors $\vc r_i$ wich can be compared with the actual position $\vc r$.
A systematic study of equation \eqref{position} for all possbile 3-combinations from the pulsar set leads to a ranking of pulsar combinations.%
\footnote{The number of 3-combinations from a set of $n$ pulsars is given by the binomial coefficient $\binom{n}{3}$. In our case $n=59$ and hence $\binom{59}{3} = 32\,509$ pulsar combinations had to be considered.} %
The averaged position errors, i.e. the arithmetic mean of the values $ |\vc r_i-\vc r|$, are plotted in Figure~\ref{posdiffs}
for the first 1000 preferred pulsar combinations. As lower limit, we obtained position errors of about 5~km, independent of the location  in the solar sytem. However, improvement in accuracy can be achieved by using pulse profile templates of higher quality.

\quad

\section{Summary and Future Prospects}

Currently, we are exploring the achievable accuracy of spacecraft navigation based on X-ray pulsar timing data and its feasibility in technical terms.
We built up a database of $\sim 60$ pulsars for which a pulsed radiation in X-rays has been detected by mid 2010 (cf. Tables 6.8 and 6.9 in Reference \mycite{becker2}). The database contains pulsar characteristics like the mean pulse profiles at various photon energies, pulsed fractions in various energy bands, pulse profile templates, i.e. approximations of pulse profiles by analytical functions and the accuracies of pulse-arrival measurements, based on least-squares fitting to the currently available pulse profile templates \mycite{prinz2}.
These data, together with the ranking of preferred pulsar combinations, serve as input to the navigation algorithm.
On the basis of simulated pulse profiles as measured by an arbitrarily moving virtual detector, we are studying parameters, such as effective detector area, sensitivity, temporal resolution, etc. As to possible implementations, weight, cost, complexity and energy budget of the system, as well as technological developments, e.g. light-weighted X-ray mirrors, have to be considered.

\end{document}